\documentclass{iaus}
\usepackage{graphicx}

\title[S266.~~B and Be Stars of h and chi Per] 
{Analysis of B and Be Star Populations of\\ the Double Cluster $h$ and $\chi$ Persei}

\author[Amber N.\ Marsh, M.\ Virginia McSwain, \& Thayne Currie]   
{Amber N.\ Marsh$^1$, M.\ Virginia McSwain$^1$
 \and Thayne Currie$^2$}

\affiliation{$^1$Lehigh University, Department of Physics, Bethlehem, PA \\ email: {\tt anm506@lehigh.edu, mcswain@lehigh.edu} \\[\affilskip]
$^2$Harvard - Smithsonian Center for Astrophysics, Cambridge, MA \\ email: {\tt tcurrie@cfa.harvard.edu}}

\pubyear{2009}
\volume{266}  
\jname{Star Clusters}
\editors{A.C. Editor, B.D. Editor \& C.E. Editor, eds.}

\include{comdef}
\begin{document}

\maketitle

\begin{abstract}
We present blue optical spectra of 92 members of $h$ and $\chi$ Per obtained with the WIYN telescope at Kitt Peak National Observatory. From these spectra, several stellar parameters were measured for the B type stars, including $V$ sin $i$, $T_{\rm eff}$, log $g_{\rm polar}$, $M_{\star}$, and $R_{\star}$. Str\"{o}mgren photometry was used to measure $T_{\rm eff}$  and log $g_{\rm polar}$ for the Be stars. We also analyze photometric data of cluster members and discuss the near-to-mid IR excesses of Be stars. 
\end{abstract}

\firstsection 
\section{Introduction}
NGC 869  and NGC 884 ($h$ and $\chi$ Persei, respectively) are a well known double cluster rich in massive B-type stars, and have been the focus of many studies over the years. Recent studies show that NGC 869 and NGC 884 have nearly identical ages of $\sim$ 13--14 Myr, common distance moduli of dM $\sim$ 11.85, and common reddenings of E(B-V) $\sim$ 0.5--0.55 (\cite[Currie et al.\ 2009]{Currie_etal09}, \cite[Slesnick et al.\ 2002]{Slesnick_etal02}, \cite[Bragg \& Kenyon 2005]{Bragg_etal05}).

\cite[Currie et al.\  (2008; hereafter C08)]{Currie_etal08} identified two populations of NGC 869 and NGC 884 stars with detected Spitzer MIPS-24 $\mu$m excess emission: 20 A and F-type stars with luminous debris disk emission and 57 brighter, earlier stars with weaker excess emission. They identify most of the latter group as candidate Be stars. However, only 21 were previously listed as Be stars (eg. \cite[Bragg \& Kenyon 2002]{Bragg_etal02}, \cite[Slesnick et al.\ 2002]{Slesnick_etal02}).

In this study, we analyze blue optical spectra of 92 early-type cluster members, including 16 candidate Be stars from C08, and investigate their near-to-mid infrared (IR) excesses. With continued monitoring of these stars in the both the optical and IR regimes, we hope to explore these excesses as a reasonable means for identifying potential Be stars within clusters, as well as to investigate the transient natures of the disks surrounding the known Be stars in NGC 869 and NGC 884. 

\vspace{-0.1in}
\section{Overview}
Blue optical spectra of 92 members of NGC 869 and NGC 884 were obtained on 2005 November 14--15 using the WIYN 3.5m telescope with the Hydra spectrograph. The observed spectra cover a wavelength range of 4250--4900 \AA.

Shown in Figure 1 are samples of the model spectral fits used to measure values for $V$ sin $i$, $T_{\rm eff}$, and log $g$ for B-type stars. $V$ sin $i$ was determined by comparing the He I $\lambda\lambda$4387, 4471, 4713, and Mg II $\lambda$4481 lines  with the Kurucz ATLAS9 models (\cite[Kurucz 1994]{Kurucz_94}) and taking a weighted average of these four values. For stars having $T_{\rm eff}$ $\ge$ 15000 K, the TLUSTY BSTAR2006 models (\cite[Lanz \& Hubeny 2007]{Lanz_07}) were used to find $T_{\rm eff}$ and log $g$ using the H$\gamma$ line. For stars having $T_{\rm eff}$ $\le$ 15000 K, the Kurucz ATLAS9 models were used (\cite[Kurucz 1994]{Kurucz_94}). The method of \cite[Huang \& Gies (2006; hereafter HG06)]{Huang_06} was used to determine log $g_{\rm polar}$. For Be stars, Str\"{o}mgren photometry  available from the WEBDA database was used to derive $T_{\rm eff}$ and log $g_{\rm polar}$ based on the methods of \cite[McSwain et al.\  (2008)]{McSwain_etal08}. The masses and radii for all stars were determined from the Schaller et al. (1992) evolutionary tracks, which are shown plotted with $T_{\rm eff}$ and log $g_{\rm polar}$ in Figure 2.  

\begin{figure}[h]
\vspace{-0.5cm}
\hspace{-1cm}
\includegraphics[scale=0.3]{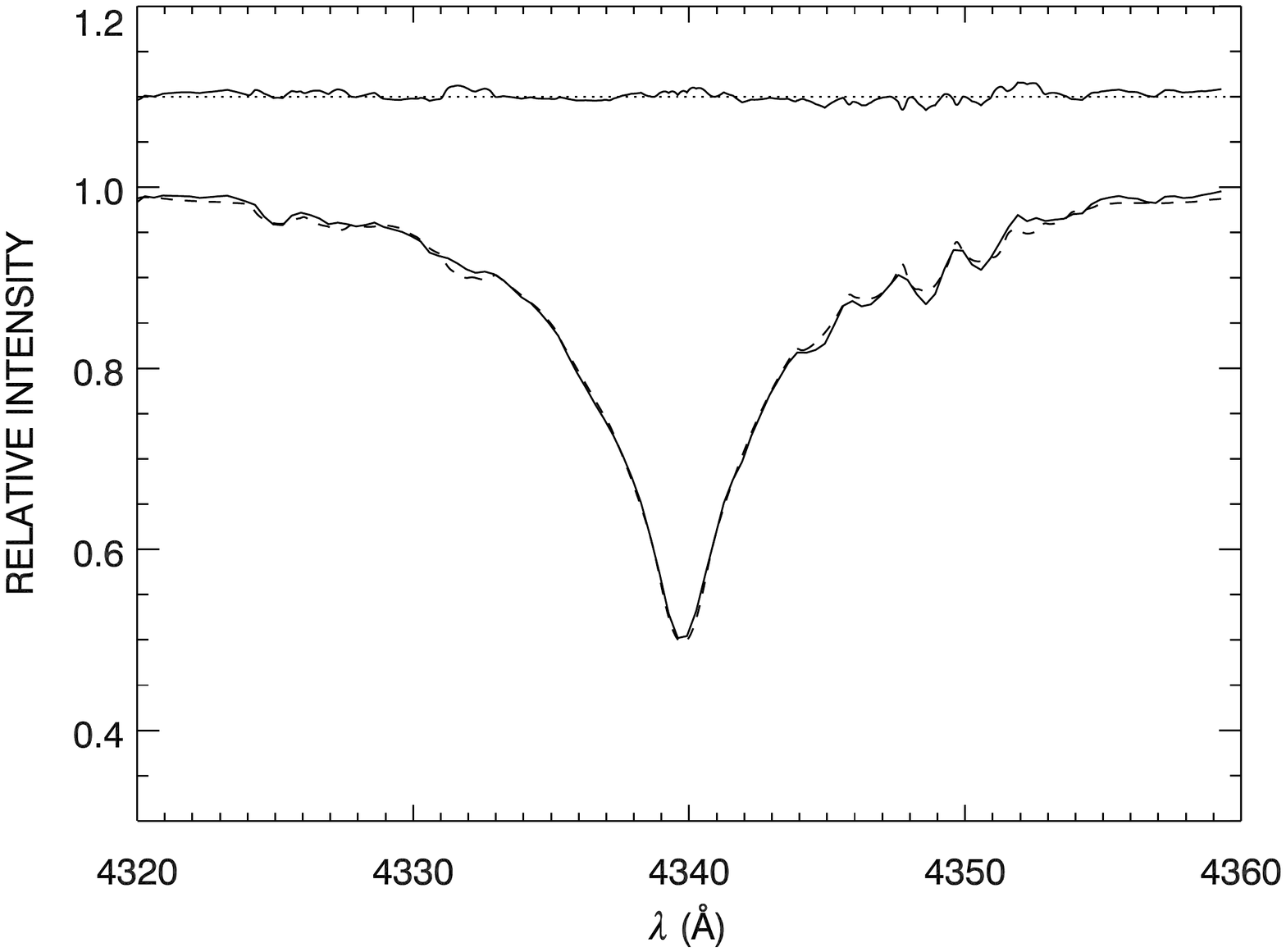}
\hspace{-2 cm}
 \includegraphics[scale=0.3]{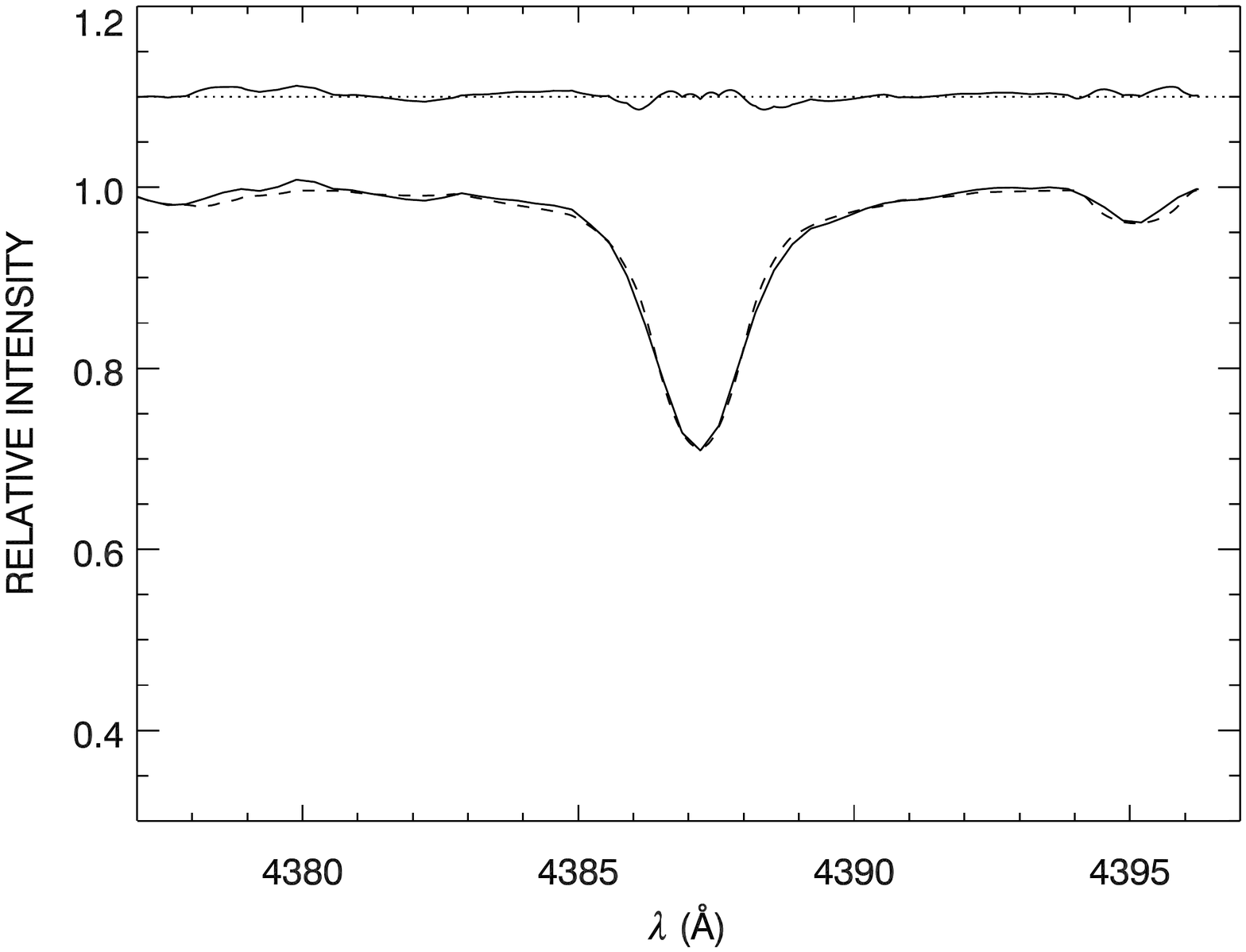} 
\caption{Sample spectral line fits for NGC 869-90. Shown on the left is H$\gamma$ and on the right is He I  $\lambda$4378. The solid line is our observed spectrum while the dashed line displays our model fit to the line, with the computed residual shown above, shifted for clarity. }
\end{figure}

\begin{figure}[h]
\vspace{-0.5cm}
\hspace{-1cm}
 \includegraphics[angle=180, scale=0.3]{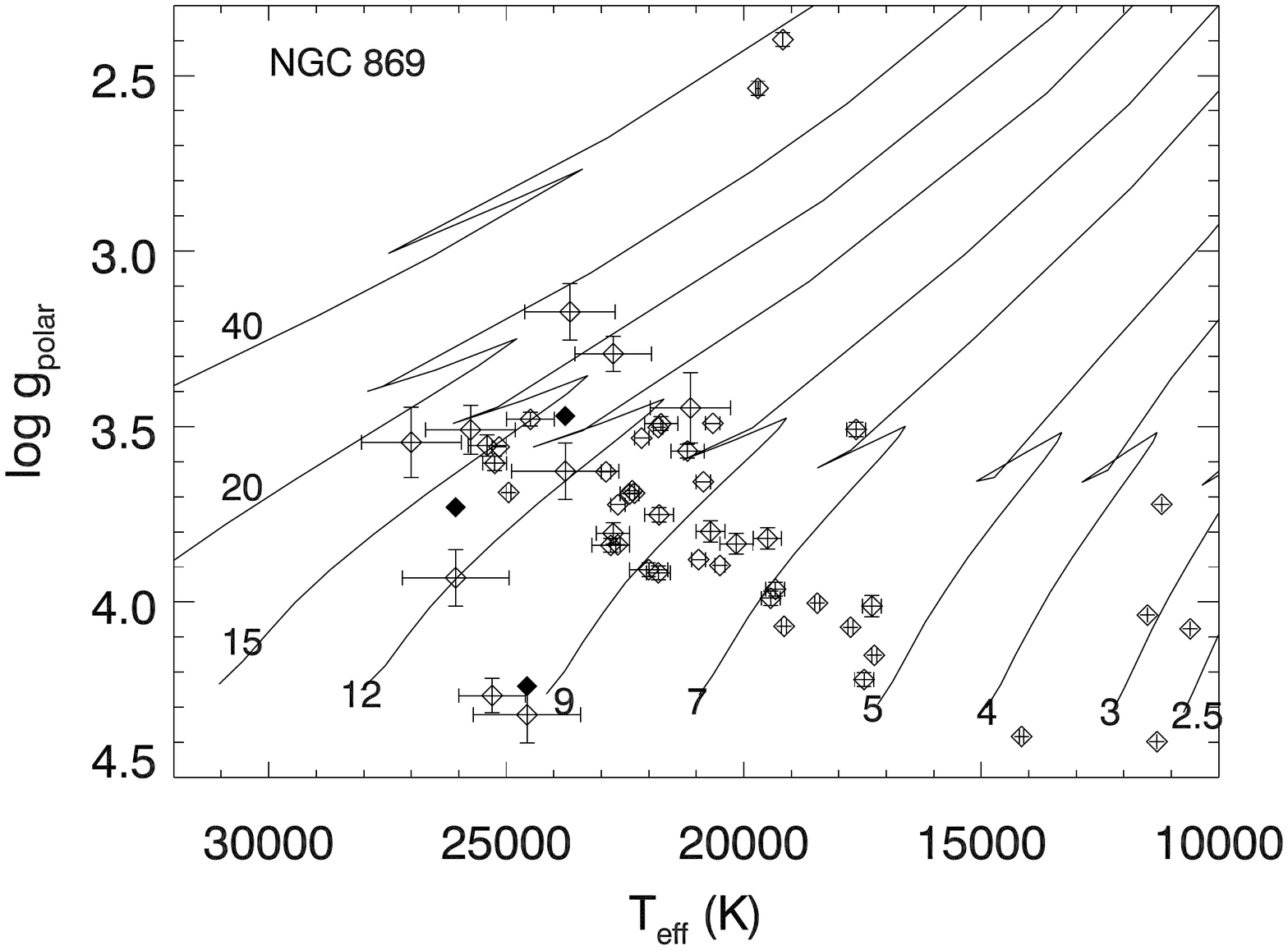} 
 \hspace{-2 cm}
  \includegraphics[angle=180, scale=0.3]{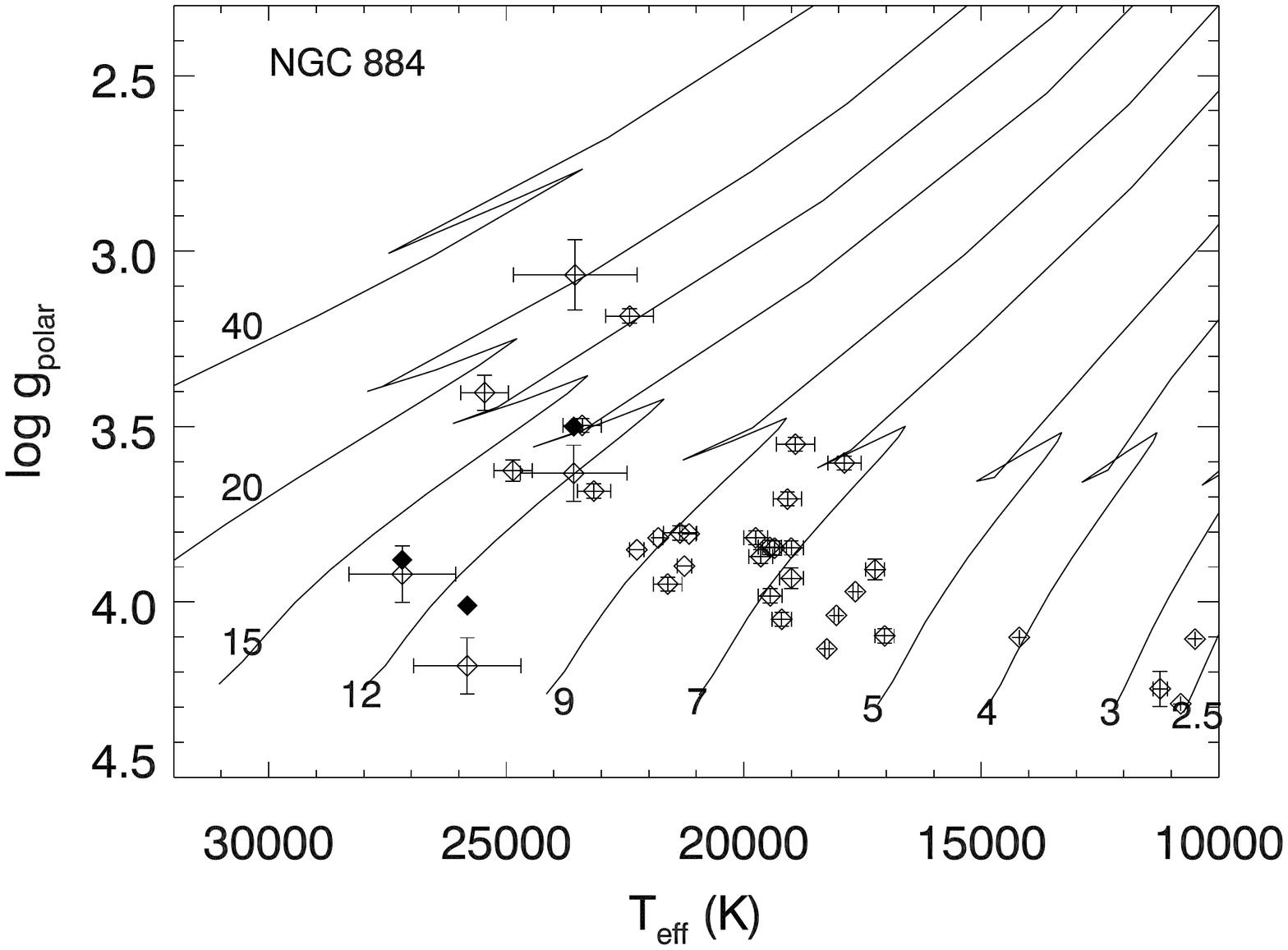} 
\caption{ For both NGC 869 (left) and NGC 884  (right), $T_{\rm eff}$ and log $g_{\rm polar}$ are plotted with the evolutionary tracks of \cite[Schaller et al.\  (1992)]{Schaller_etal92}. The ZAMS mass of each evolutionary track is labeled along the bottom. Normal B-type stars are shown as open diamonds while Be stars are filled diamonds. }
\end{figure}

\section{Results}
Sixteen Be candidates from C08 are present in our sample or that of HG06. Three of these 16 stars show no evidence of circumstellar emission in our spectra, though all have been observed to be Be stars in the past (\cite[Keller et al.\ 2001]{Keller_etal01}). Ten of the C08 Be candidates in our spectra do show emission. Stellar parameters for the remaining three candidates are found in HG06, thus we cannot comment on the presence of emission at the time of observation. In addition, we find Be emission in one star (No. 1772) that was not observed by C08, and we present results for one additional star (No. 1268) identified as a Be star by \cite[Keller et al.\ (2001)]{Keller_etal01}. These results are summarized in Table 1. 

\begin{table}
 \begin{center}
 \caption{Measured physical parameters for Be stars}
 \label{table1}
 \scriptsize
 \begin{tabular}{lccccccll}\hline
   {\bf Cluster-} & {\bf $V$ sin $i$} & {\bf $T_{\rm eff}$} &{\bf log $g_{\rm polar}$} & {\bf $M_{\star}$} & {\bf $R_{\star}$} & {\bf Be} & & \\
   {\bf Star$^1$} & {\bf (km~s$^{-1}$)} & {\bf (K)} & {\bf (dex)} & {\bf ($M_{\odot}$)} & {\bf ($R_{\odot}$)} & {\bf Cand.$^2$}& {\bf Comment} & {\bf Ref.} \\ \hline
   NGC 869-49 & 172 & 23757 & 3.63 & 12.3 & 8.9 & Y & Emission present  & This work\\
   NGC 869-517 & 178 & - & - & - & - & Y& Emission present & This work\\
   NGC 869-566 & 306 & 21183 & 3.57 & 10.4 & 8.7 & Y & No emission present$^3$ & This work\\
   NGC 869-846 & 205 & 22747 & 3.29 & 14.7 & 14.3 & Y & Weak emission present & This work\\
   NGC 869-847 & 87 & 27000 & 3.54 & 17.8 & 11.8 & Y & Weak emission present & This work\\
   NGC 869-1162 & 66 & 19175 & 2.40 & 33.6 & 60.8 & Y & No emission present$^3$& This work\\
   NGC 869-1261 & 285 & 26065 & 3.93 & 12.0 & 6.2 & Y & Strong emission present & This work\\
   NGC 869-1268 & 151 & 24491 & 3.48 & 14.8 & 11.6 & N & No emission present$^{3, 4}$ & This work\\
   NGC 869-1278 & 197 & 24562 & 4.32 & 9.0 & 3.4 & Y & Emission present & This work\\
   NGC 884-1772 & 379 & - & - & - & - & -  & Emission present  & This work\\
   NGC 884-1926 & 106 & 27190 & 3.92 & 13.5 & 6.7 & Y & Strong emission present & This work\\
   NGC 884-2091 & 236 & - & - & - & - & Y & Emission present & This work\\
   NGC 884-2138 & 153 & 23579 & 3.63 & 12.0 & 8.7 & Y & Emission present & This work\\
   NGC 884-2165 & 79 & 26571 & 4.03 & 11.9 & 5.5 & Y & -- & HG06 \\
   NGC 884-2402 & 141 & 28238 & 3.81 & 15.6 & 8.1 & Y & -- & HG06 \\
   NGC 884-2468 & 134 & 10500 & 4.11 & 2.7 & 2.4 & Y & No emission present$^3$& This work\\
   NGC 884-2563 & 308 & 25820 & 4.18 & 10.7 &  4.4 & Y & Strong emission present & This work\\
   NGC 884-2949 & 168 & 18240 & 3.96 & 6.4 & 4.4 & Y & -- & HG06 \\ \hline
 \end{tabular}
 \end{center}
 \vspace{1mm}
 \scriptsize
 {\it Notes:}\\
 $^1$ Identification numbers from the WEBDA database.
 $^2$  Be candidate in C08.
 $^3$ Stars not showing emission in our observations are likely transient Be stars.
 $^4$ Identified as Be star by \cite [Keller et al.\ (2001)]{Keller_etal01}.
\end{table} 

\begin{figure}[b!]
\begin{center}
\includegraphics[angle=180,width=3.in]{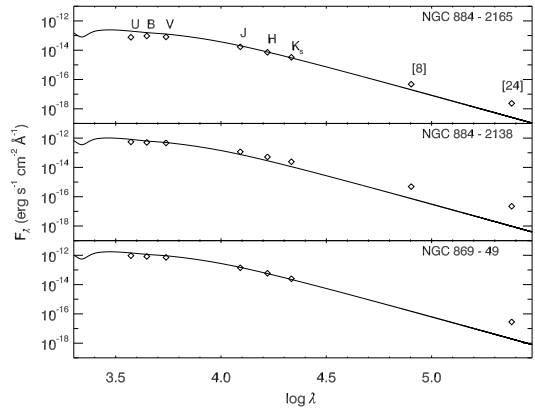} 
\caption{SEDs for three stars in NGC 869 and NGC 884. Reddened blackbody curves are overlaid with these plots to investigate their near-to-mid IR excesses.}
\end{center}
\label{sed}
\end{figure}

 Spectral energy distributions (SEDs) for three stars in NGC 869 and NGC 884 are displayed in Figure 3. UBV magnitudes are  from the WEBDA database, JHK$_{s}$ are from the 2MASS survey, and Spitzer [8] and [24] $\mu$m are from C08. These magnitudes were then converted to fluxes via the methods detailed in \cite[Bessell et al. (1998)]{Bessell_etal98}, \cite[Cohen et al.\ (2003)]{Cohen_etal03}, \cite[Colina et al.\ (1996)]{Colina_etal96}, \cite[Reach et al.\ (2005)]{Reach_etal05}, and  \cite[Rieke et al.\ (2008)]{Rieke_etal08}. Assuming a constant $E(B-V)$ = 0.52 for NGC 869 and NGC 884 (\cite[Bragg \& Kenyon 2005]{Bragg_etal05}, \cite[Slesnick et al.\ 2002]{Slesnick_etal02}), reddened blackbody curves have been overlaid with these plots to investigate their near-to-mid IR excesses. All three stars shown in Figure 3 are proposed Be candidates (C08), with NGC 884-2138 and NGC 869-49 having emission present in our optical spectra and observed near-to-mid IR excess. NGC 884-2165 is not in our spectroscopic sample but has previously been identified as a Be star and has observed IR excess (\cite[Keller et al.\ 2001]{Keller_etal01}).

\section{Conclusions and Further Work}
We have measured the physical parameters of 77 B-type stars and 15 Be stars in NGC 896 and NGC 884. Sixteen Be candidates from C08 are present in our sample or that of HG06. Of these 16 Be candidates, 3 stars show no evidence of emission in our optical data and are likely transient Be stars. Ten of these Be candidates do show emission in our spectra. Those Be candidates without emission in our spectra should be monitored in the future to further investigate their transient nature. 

In the future, IRAC 3.6-5.8 $\mu$m data will be combined with the optical and IR fluxes used here to investigate the observed SEDs. We will fit the new SEDs using modern flux models rather than blackbody curves. Modifications accounting for variable reddening throughout the clusters will also be made. These new SED fits can then be used to model the Be disk sizes and temperatures. 

\acknowledgments
We are grateful for travel support provided by the American Astronomical Society and the International Astronomical Union. We would like to thank Yale University for providing access to the WIYN telescope at KPNO. Institutional support was provided by Lehigh University. This work was also supported by NASA DPR number NNX08AV70G.  


\end{document}